# CROSSCON: Cross-platform Open Security Stack for Connected Devices

Bruno Crispo, Marco Roveri (Uni. Trento); Sandro Pinto, Tiago Gomes (Uni. Minho); Aljosa Pasic (ATOS); Akos Milankovich (S-LAB); David Purón, Ainara Garcia, (Barbara IoT); Ziga Putrle, (Beyond Semiconductor); Peter Ten (Uni Wuerzburg); Malvina Catalano (Cysec)

*Abstract*—The proliferation of Internet of Things (IoT) embedded devices is expected to reach 30 billion by 2030, creating a dynamic landscape where diverse devices must coexist. This presents challenges due to the rapid expansion of different architectures and platforms. Addressing these challenges requires a unified solution capable of accommodating various devices while offering a broad range of services to connect them to the Internet effectively. This white paper introduces CROSSCON, a three-year Research and Innovation Action funded under Horizon Europe. CROSSCON aims to tackle current IoT challenges by developing a new open, modular, and universally compatible IoT security stack. This stack is designed to be highly portable and vendor-independent, enabling its deployment across different devices with heterogeneous embedded hardware architectures, including ARM and RISC-V. The CROSSCON consortium consists of 11 partners spanning 8 European countries. This consortium includes 4 academic institutions, 1 major industrial partner, and 5 small to medium-sized enterprises (SMEs).

## I. INTRODUCTION

CONNECTED devices are becoming increasingly complex [1], resulting in additional deployment challenges and new security concerns. This especially holds for "things", which is how usually small and low-end connected devices are called. Examples of Internet of things (IoT) devices include general sensors and actuators, and appliances that connect wirelessly to a network and can transmit data. Different stakeholders have different perspectives on IoT security challenges and concerns [2]:

- **Manufacturers:** focus on enhancing assurance and security levels in their designs while simultaneously reducing development and production costs;
- **Embedded and IoT application developers:** focus on implementing the required functionality of the end system, while keeping the security and low-cost aspects;
- **Policy makers and regulators:** play a crucial role in assessing the societal, legal, and economic impacts of cybersecurity in systems composed of connected devices;
- **Standardization and certification bodies:** prioritize flexible and modular schemes that align with regulatory actions while considering market realities.

To address these pressing challenges, the research project ***CROSSCON: Cross-platform Open Security Stack for Connected Devices*** was launched in 2022 with support from the Horizon Europe programme. The CROSSCON consortium comprises 11 partners from 8 European countries: Spain, Italy, Portugal, Slovenia, Hungary, Germany, Poland, and Switzerland. This consortium includes a major industrial partner, Eviden, along with 4 academic institutions - the University of Trento, the University of Minho, the University of Würzburg, and the Technical University of Darmstadt. Additionally, there are 5 Small and Medium-sized Enterprises (SMEs) involved: Beyond Semiconductors, CYSEC, 3MDEB, Search Lab, and Barbara IoT. Together, these partners bring a wealth of technological and scientific expertise, industrial and end-user perspectives, as well as valuable business and market insights.

With this white paper, we aim to provide: (i) a brief overview of the current IoT landscape; (ii) insights into the current challenges and motivations driving the development of an IoT security stack; and (iii) details about the CROSSCON's technical approach and the use cases that will be implemented to validate the project's contributions.

## II. CROSSCON IN A NUTSHELL

There are many security risks associated with IoT devices [3], [4], which include expanded attack surfaces, unsecured hardware, inadequate IoT lifecycle management, and firmware exploits. Tackling these risks poses several challenges due to the diverse range of devices, each with its own unique security features and capabilities. Furthermore, the complexity escalates when devices lacking security-related hardware features are deployed in security-critical contexts, such as healthcare, critical infrastructure, or automotive systems, making them attractive targets for cyber-attacks.

In this very heterogeneous landscape, IoT devices vary greatly in terms of hardware, ranging from (i) bare-metal devices featuring low-power microcontrollers (MCUs) with few kilobytes of RAM and tens of kilobytes of Flash memory,

First version of the white paper: April, 2024; date of current version April, 2024. This work is supported by the European Union's Horizon Europe research and innovation program under grant agreement No 101070537, project CROSSCON (Cross-platform Open Security Stack for Connected Devices).

to (ii) devices equipped with powerful application processing units (APUs) boasting multiple cores, and even to (iii) reconfigurable hardware devices with customizable logic gates or domain-specific hardware architectures for efficient implementation of high-complexity tasks. However, this wide diversity expands to the security features present in both the hardware and the firmware [5], [6], which can also range from zero hardware-enabled security to the implementation of Root of Trust (RoT) and cryptographic engines.

A typical IoT infrastructure comprises several layers, including hardware, firmware, the operating system (OS), network stack, cloud infrastructure, and more, each contributing to the overall attack surface. It's crucial to recognize that the security of a system is only as strong as its weakest component. Any IoT device lacking security assurances can serve as a Trojan horse, enabling attackers to infiltrate the IoT system and laterally move to other security-sensitive targets. This discrepancy highlights a significant misalignment of incentives between device manufacturers, who focus on the initial product design or foundational trusted services, and IoT software developers, who must contend with the ongoing and dynamic processes inherent in IoT environments.

Devices rely on foundational trusted services, such as secure boot, and additional trusted services like update and patching mechanisms, as well as software component isolation. Patching and upgrades require coordination among various organizations in the value chain. Policymakers, regulators, standardization, and certification bodies have taken note of these challenges. For instance, the European Commission proposed the Cyber Resilience Act, advocating for mandatory security patches for connected devices capable of collecting and sharing data. However, a deeper understanding of diverse perspectives and the impact across the entire IoT system value chain is needed, along with strategies for implementing this emerging "secure connected device" scenario.

The CROSSCON stack aims at addressing these challenges by providing a new open, flexible, highly portable, and vendor-independent IoT security solution that can operate across various hardware platforms. This approach aligns with the vision of establishing trustworthy IoT devices across the IoT infrastructure. The CROSSCON security stack aims to enable interoperability of hardware security features and essential security services, such as attestation, secure boot, and update mechanisms, across different platforms. This stack will be directly usable by Hardware Manufacturers, Original Device Manufacturers (ODMs), Application Developers, and System Integrators, enabling them to build devices and applications with fewer resources, eliminate hardware dependencies, and ensure a consistent and robust security foundation across the entire IoT ecosystem. The CROSSCON IoT security stack will run on various edge devices and computing architectures, including those built on the open RISC-V ISA.

### III. SUPPLY AND VALUE CHAIN FOR IOT DEVICES

There are several security challenges currently being faced by the IoT industry, from both the consumer and the services provider sides, affecting different actors, such as the Hardware Manufacturers, Original Device Manufactures (ODMs), Application Developers, and System Integrators. While the Hardware Manufacturers ares responsible for the design and production of the physical components of a device, which can include sensors, MCUs, communication modules, and other relevant hardware, ODMs focus on the design, manufacture, and the delivery of the device. The ODMs not only assemble the final device hardware, but also develop and integrate the embedded software (i.e., the firmware) for basic functions such as device boot process, sensor data collection, and communication establishment. Regarding the application Developers, they are responsible to implement the software application running on the IoT devices (that are provided by the ODM) that brings value to device's end-users and business. Finally, the System Integrators are the companies responsible to integrate connected devices into existing infrastructure, ensuring that they can seamlessly communicate with each other and with server platforms.

#### A. Security challenges: Service providers

*Hardware Manufacturers:* Hardware manufacturers primarily contend with interoperability and standardization challenges. Ensuring that hardware components are compatible with other products while adhering to security standards can be complex, particularly in heterogeneous environments like the connected device landscape. Additionally, they must address side-channel attacks, such as power analysis, electromagnetic analysis, and timing attacks, which can be exploited to extract sensitive data from devices. Consequently, manufacturers need to implement increasingly intricate countermeasures to mitigate these risks.

*ODMs:* The IoT devices designed by the ODM are susceptible to firmware and hardware vulnerabilities, requiring for regular attention and patching through Over The Air (OTA) updates. This represents a challenge with high impact, as hardware often has a lengthy life-cycle, and maintaining security over extended periods can be resource-intensive. Implementing device Secure Boot and Chain of Trust (CoT), which serve as foundational trusted services to prevent unauthorized code execution and device tampering, also poses a significant challenge, as some of these foundational trusted services might rely on hardware-enabled security features.

*Application Developers:* Application developers usually face interoperability challenges, especially when enhancing their application security capabilities. Older or more basic hardware architectures may lack inherent security features and are difficult to retrofit. Even with modern architectures, there is no standardization of the security stack among different IoT platforms and devices, while some application services might use or depend upon foundational trusted services provided by the ODM.

*System Integrators:* System integrators find it imperative to implement robust authentication protocols, encryption mechanisms, and access control systems. These measures are crucial to ensure the integrity and security of the systems they integrate. In addition, security mechanisms also rely on the existing building blocks provided by ODM and hardware manufacturers, creating another link in the Chain of Trust.

### B. Security challenges: Affected Industries

While ensuring security in consumer devices presents several challenges, some industries, due to the nature of their activities, can face additional security concerns.

*Healthcare:* The healthcare industry heavily relies on connected devices for patient monitoring, medical equipment, and electronic health records. Ensuring the security and privacy of patient data and the reliability of medical devices is of paramount importance.

*Industrial and Manufacturing:* Industries employing IoT devices for industrial automation, process control, and predictive maintenance are deeply concerned about security. Any breach in these systems can result in production downtime, damage, or even safety risks.

*Energy and Utilities:* The energy and utility sectors make extensive use of IoT devices for grid management, smart meters, and energy distribution. Security is essential to safeguard critical infrastructure and to prevent service disruptions.

*Automotive:* The automotive industry is progressively becoming more connected, with vehicles incorporating numerous IoT devices for navigation, infotainment, and safety features. Security is of utmost importance to prevent vehicle hacking and to ensure passenger safety.

### C. CROSSCON Contributions

With all this in mind, the CROSSCON security stack can bring additional value by addressing the challenges mentioned before, as it could provide developers with the tools and capabilities required to elevate level of assurance for these security services. In its turn, implementation of these services and CoT not only addresses significant technical challenges prevalent in the aforementioned industries, strengthening security measures and mitigating potential risks, but also responds to another emerging requirement for IoT devices, platforms and infrastructures, which is related to the compliance with important EU regulations and relevant certifications.

Hardware Manufacturers and Application Developers can benefit from CROSSCON by gaining access to a standardized security framework that streamlines interoperability, ensuring their products are compatible across devices. Additionally, the built-in security features and services developed as part of CROSSCON can serve as a defense against the advanced attacks mentioned above, posing a great opportunity for these Hardware Manufacturers and System Integrators to simplify the implementation of otherwise very complex security measures. CROSSCON will also provide ODMs and Application Developers the foundational enabling capabilities to build seamless updates OTA, enabling secure patching and vulnerability addressing of their firmware and software.

### IV. EU POLICY AND REGULATION

There are several legislative areas related to CROSSCON, such as the NIS2 Directive[1] (Directive EU 2022/2555), the Cybersecurity Act[2], which establishes a cybersecurity certification framework for products and services, and the proposed Cyber Resilience Act (CRA)[3]. The European Chips Act[4] links to some of these EU certification schemes (article 17), while NIS2 provisions about ICT supply chain security also invite stakeholders to identify *"the specific services, systems or products that might be subjected to the coordinated supply chain risk assessments with priority"*. NIS1 used to target *"operators of essential services"*, which are entities working in the 7 critical infrastructure sectors: energy; transport; banking; financial market infrastructures; health; drinking water supply; and digital infrastructure, and *"digital services providers"*, which correspond to digital platforms such as cloud, search, and e-commerce. However, NIS2 targets many other sectors of high criticality including public administrations, manufacturing, or postal services (*"essential entities"* and *"important entities"*). By 17 October 2024, Member States must adopt and publish the measures necessary to comply with the NIS2 Directive, which apply those measures from 18 October 2024. The measures include provisions relevant to CROSSCON, such as supply chain security, including direct suppliers or service providers of IoT devices.

The most relevant regulation for CROSSCON, however, is the CRA, which is imposing several obligations for manufacturers, distributors, and importers of connected devices. It is also addressing cybersecurity essential requirements across the life-cycle of these devices, standards to follow, and conformity assessment procedures. CRA is a legal framework that describes the cybersecurity requirements for hardware and software products placed on the market by manufactures that will be obliged to manage security throughout a product's life-cycle. In annex III (reference) microcontrollers, for example, are placed in the class I of critical devices, while CPUs are examples of products in class II, together with operating systems or secure elements. Both CROSSCON and CRA advocate for **"security by design"**, meaning that cybersecurity will have to be considered from design and development phase, sing when possible, hardware-enabled security and foundational trusted services to elevate level of assurance. The so called "trusted services", some of which are tested in CROSSCON use cases, are commonly used in operation (e.g., multi-factor authentication), delivery (e.g., commissioning of IoT device) and maintenance (e.g., firmware update) phases. Although the open-source CROSSCON secure stack is non-commercial and as such it is not subject to CRA, it can be used in commercial solutions which are subject to it.

Some stakeholder groups already expressed their opinion on CRA, through their associations. Alliance for Internet of Things Innovation (AIOTI), for example, is the multi-stakeholder platform for IoT Innovation in Europe. In their paper on Impact Assessment on Cyber Resilience Act [7] they mention life-cycle approach to cybersecurity in IoT, which does not imply that one sole party such as an upstream manufacturer, midstream integrator or downstream customer is responsible for the whole device's life-cycle. They discuss that the CRA is aiming to establish for each actor their roles, (co)responsibilities and related (co)accountability to ensure legal certainty. They also request ecosystem thinking, as well as the harmonization in different critical, vital or essential IoT ecosystems. European Cyber Security Organisation (ECSO) issued a Technical Paper on Internet of Things (IoT) in July 2022 [8] that aims to identify current and foreseen challenges related to IoT cybersecurity at technical level, at regulatory level, and in relation to certification.

---

[1] https://eur-lex.europa.eu/eli/dir/2022/2555
[2] https://digital-strategy.ec.europa.eu/en/policies/cybersecurity-act
[3] https://digital-strategy.ec.europa.eu/en/library/cyber-resilience-act
[4] https://commission.europa.eu/strategy-and-policy/priorities-2019-2024/europe-fit-digital-age/european-chips-act_en

The ECSO Working Group 1 and Policy Task Force also issued position paper on CRA which was updated in 2023. Besides alignment and harmonisation, ECSO mentions that it is essential for companies to have a clear methodology for risk assessment and product categorization, knowing in advance whether their products will fall under the default category, Class I or Class II, so that they can adjust their internal processes and invest for the right conformity assessment methods. ECSO mentions that IoT devices are deployed in the field provide a large attack surface, as they are also susceptible to physical tampering and attacks. According to their technical paper, trust starts with the processors at the core of each IoT device to create trustworthy platforms. They further mention RISC-V and consider European sovereignty over the implemented circuits to facilitate modifications at the core of the architecture and tackle device, sector, or application-specific security constraints.

Another challenge that is mentioned by ECSO is the device's life-cycle management, since there are devices designed for a short life span, while there are others that must be alive for decades, which raises issue of firmware and application integrity, and delivery updates. Scalable remote attestation procedures are needed, while protection against advanced physical attacks such as side-channel and fault attacks for data and intellectual property protection are also high on their list of challenges. They also mention protection against micro-architectural attacks such as Spectre [9] or Meltdown [10] in devices with low computing power constraints. CROSSCON is especially relevant in design and development phase, both for the assessment of the risks associated with a product, as well as product-related essential requirements (Annex I, Section 1 of CRA). Conformity assessment, which is addressed in Annex IV, could also be relevant. In this line there is Draft Standardisation Request in support of the CRA, where CROSSCON could also contribute. Preparatory work on standards already started and the proposed approach is to build on work done for RED DA (Delegated Act of the Radio Equipment Directive (EU) 2022/30 that was already adopted by the European Commission). Preliminary draft identified 13 "horizontal" standards for CRA essential requirements and 30 "vertical" standards for critical products, to be defined by May 2025 and May 2026, respectively.

CROSSCON is monitoring these developments and drafting which is done iteratively with all stakeholders. It is involved in public consultation, and contributions through different bodies (e.g., ECSO). CROSSCON could contribute to CRA discussions related to obligations for manufacturers to provide security updates for the entire product life cycle. While CROSSCON is addressing secure updates, not exclusively limited to security updates, we also underline the fact that any updates, patching, and especially security updates, also need to have guarantee that they are done in a secure manner.

An insecure update process, whether it is firmware or security patch, also presents a major issue as it allows an attacker to upload malicious code on the device. However, problem of secure updates persists, since: (i) updates often come as a bundle of code snippets developed by different parties; and (ii) code might be signed by digital signatures, but these do not give any guarantee on how the functionality of the device changes by the update. Furthermore, CROSSCON considers two types of updates:

- **Full update**: the package contains the full replacement of the old package to be installed.
- **Partial update:** the package contains just the binary difference or a software patch version between the new firmware and the old version.

Finally, besides CRA, there is also a clear link to the revision of Regulation (EU) 910/2014, known as eIDAS2 Regulation *"electronic Identification, Authentication, and trust Services"*. This is an EU regulatory framework that establishes a set of rules and standards for eID and trust services in the member countries of the European Union and it promotes the European Digital Identity Wallet (EUDI Wallet) as an app that enables citizens and residents all over the EU to identify and authenticate themselves. The highest level of assurance requires secure hardware solutions for the security and strength of the Wallet. CROSSCON could contribute by enhancing security of EUDI since instead of "black box" RoT, CROSSCON is enabling own implementation of RoT inside the TEE.

## V. CROSSCON Technical Approach

The security services that are commonly required in IoT devices and platforms include (but are not limited to):

- **Cryptography Services and Secure Communications:** to support encryption, decryption, digital signatures, and other cryptographic functions intended to establish secure communication channels with remote servers or other trusted entities, ensuring the confidentiality and integrity of data in transit;
- **Secure Storage for sensitive data**: such as encryption keys and authentication tokens. This data should be protected from being accessed or exfiltrated by unauthorized applications or malicious actors;
- **Authentication and Identity Management:** to manage securely device identity and authentication credentials, enabling secure authentication and identity management between devices and with server applications;
- **Remote Attestation:** to attest to device's integrity and security state, allowing remote parties to verify that the device is operating as expected;

- **Access Control:** that ensures only authorized applications or users can interact with secure services and data within the device;
- **Secure Boot:** that ensures that only authorized software is able to boot, more precisely that the device starts in a known (secure) state;
- **Secure Updates:** to ensure that if there is an update, it can be securely applied, protections against downgrading could also be built in.

Adversaries can exploit software vulnerabilities to take control of a device remotely or can perform local attacks such as side-channel and fault attacks on memories. Countermeasures that are rooted in the hardware are needed, in order to ensure efficiency and strong security guarantees. This is where RoT come into play. These RoT are hardware, firmware, and software components that perform specific, critical security functions. They are inherently trusted, secure by design, and a starting point for the CoT. IoT platforms with hardware-enabled security are commonly used to provide fixed interfaces for hardware-protected execution environments. Typically, functionality cannot be easily added or scaled according to the IoT application needs. There is a lack of a flexible and open software framework based on open-source hardware. TPMs (Trusted Platform Modules) are an example of technology that can be used to separate critical functionality from more vulnerable user applications and contains countermeasures against side-channel and fault attacks. However, developing own security applications for TPMs is not foreseen and they are seen as "black boxes", as they have pre-defined functionality. On the other hand, hardware mechanisms for establishing compartmentalization via TEEs also allow achieving efficient and strong isolation of software but are more flexible than TPMs.

The problem is exacerbated by the fact that two instances of the same class of device may use different hardware (i.e., ARM vs RISC-V), each implementing somehow proprietary instances of the security features and mechanisms (i.e., ARM OP-TEE, Intel SGX, Trustonic TEE, Qualcomm TEE, RISC-V Keystone, etc.) that prevent or make it very difficult in practice for the security services of the two devices to interoperate for an operating system or an application to use seamlessly the trusted services of both devices. Developers must trust the device manufacturer to have built the TEE correctly, so that attestation is valid. TEE security can be compromised through, if isolation is not implemented, allowing, for example, access to cryptographic keys or sensitive user data. In addition, hardened CPU and memory isolation that many platforms already offer is still not enough to guarantee full isolation as many micro-architectural resources such as last-level caches, interconnects, network stacks and memory controllers remains shared among partitions and processes.

Existing TEE implementations are under scrutiny, on the way they implement isolation and memory protection. There are several reported vulnerabilities[5] [3], [11], [12] that violate memory isolation when trusted applications inside the TEE interact with non-trusted ones, highlighting a more general issue with side-channel attacks. Thus, it is necessary to also investigate and find solutions to the problem of side-channel attacks looking at all the spectrum of available TEEs. All existing TEEs come with a minimal set of foundational trusted services (i.e., secure boot, secure storage, basic cryptographic functions, secure attestation, etc.) used as building blocks to implement security at the higher levels. As devices are getting more powerful and use cases become more complex, the need to improve and enrich these foundational trusted services to support the above operational security issues of IoT systems is emerging.

[5]CVE-2015-6639: https://nvd.nist.gov/vuln/detail/cve-2015-6639

CROSSCON is also exploring a two-fold opportunity related to the choice of open source hardware RISC-V. Firstly, we believe we can contribute and help shape ongoing Trusted Execution and Confidential Computing activities and specifications. Secondly, RISC-V offers a unique opportunity to provide more robust security guarantees, by enabling unseen freedom to implement novel extensions and mechanisms at the hardware level [13]. Besides addressing stakeholders needs, such as the availability of an innovative IoT open-source security stack or offering of a set of novel and high assurance trusted services, CROSSCON is also having several technical objectives, such as:

- Strengthen memory protection and isolation in new and existing TEEs, mitigating the impact of side-channel attacks.
- Provide methodology, techniques, and related tools to formally verify "correct by design" secure open-source software and firmware for connected devices.
- Provide a toolchain that integrates and validates lightweight techniques for security assurance.
- Provide validation and testing methodology, a replicable testbed, and testing and validation results for CROSSCON innovations.

To achieve such objectives, CROSSCON specifies a novel security architecture, as shown in Figure 1. The purpose of the CROSSCON architecture is to enable the secure and isolated execution of security-sensitive tasks on a wide variety of IoT devices having very different hardware security primitives. The goal of CROSSCON, therefore, is to define a flexible and adaptable set of architectural components that can provide a set of security features that maximize the use of the underlying hardware platform capabilities to provide the best possible level of isolation for sensitive workloads. The CROSSCON secure stack includes several components embedding novel technologies developed and/or extended during the project. In particular, an essential component of the stack is the CROSSCON Hypervisor. The hypervisor's ultimate objective is to create and support distinct and isolated virtual machines (VMs) that can act as virtual TEEs or enclaves, ensuring they run as if they were operating independently on separate hardware [14], [15].

The hypervisor operates within a dedicated layer, with higher privileges than the operating system, safeguarding hardware resources and leveraging various isolation mechanisms. The CROSSCON Hypervisor guarantees, by design, that hardware resources are not shared across different execution environments (in the form of VMs) and provide a set of built-in mechanisms (e.g., cache coloring) to guarantee strong isolation not only at the architectural but also micro-architectural level. The hypervisor will ensure the correct enforcement of the access control policies to guarantee that VMs can securely execute security-sensitive workloads, for example, running a Trusted operating system (OS) and Trusted Applications. The project will validate its approach in use cases that implement trusted services for connected devices, including patches, security audits, commissioning and decommissioning, and secure authentication and communications.

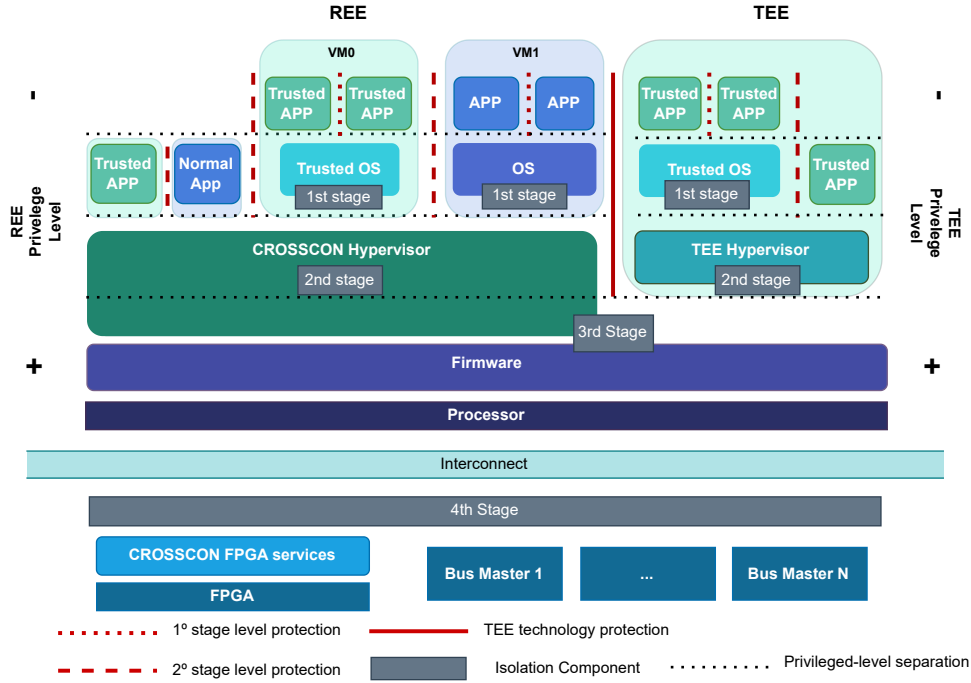

Fig. 1: Overview of the CROSSCON security stack.

## VI. THE ROLE OF CERTIFICATION

In the dynamic world of the IoT, ensuring the interoperability, security, and reliability of billions of interconnected devices is essential. This is where certification and standardization play an indispensable role. These mechanisms not only ensure the quality and security of IoT devices but also foster trust among consumers and stakeholders. In the European context, the EUCC (European Cybersecurity Certification Scheme on Common Criteria) emerges as a key certification. This scheme targets ICT products, encompassing both hardware and software products and components. Developed by ENISA with the support of an Ad-Hoc Working Group and Member States, the candidate scheme received a positive opinion from the ECCG (The European Cybersecurity Certification Group). Following this, the European Commission is set to transform it into an Implementing Act, marking its official entry into force. CROSSCON closely follows the developments and guidelines of the EUCC scheme.

Examining IoT-specific certifications, ETSI EN 303 645 standard addresses the cybersecurity requirements of consumer IoT devices. Organizations like the IoXt Alliance and the IoT Security Foundation (IoTSF) are at the forefront of enhancing security across IoT devices. The GSMA IoT guidelines cater to IoT services in the mobile industry, and the OWASP IoT project highlights top security concerns specific to IoT. The Eurosmart IoT SCS further underscores the importance of safeguarding IoT devices against cyber threats. Fixed-time certifications (e.g. EN 17640[6]) also arise to solve the issue related to duration and cost of the existing certifications, such as Common Criteria, that are not suitable for low assurance products. CROSSCON takes these certifications into account during the security requirement definitions and evaluations of the CROSSCON stack.

Apart from more generic IoT certification, TEE-related certifications and standards are even more relevant for CROSSCON. GlobalPlatforms, known for its focus on secure and interoperable digital services and devices, has introduced the SESIP approach. This approach focuses specifically on the core features and functionalities of IoT devices, simplifying the certification process by incorporating certifications for individual components. By adhering to SESIP, manufacturers and developers can ensure their IoT solutions align with the highest security benchmarks, instilling confidence in end-users. PSA by ARM is another widely accepted certification in the industry with more than 80 different vendors manufacturing certified devices. The approach involves making security-by-design based on root of trust, attestation, secure boot, isolation, secure update, storage and cryptography. CROSSCON uses the GlobalPlatforms APIs to provide compatible and secure solutions and examines the PSA certification re-quirements to provide a secure stack. As the IoT ecosystem continues to expand into various facets of our daily lives, the role of certification and standardization becomes increasingly significant. These frameworks not only set the bar for device quality and security but also cultivate a sense of trust among consumers and stakeholders. By adhering to recognized standards and obtaining relevant certifications, the industry can ensure a safer, more reliable, and interconnected digital future.

[6]https://www.cencenelec.eu/news-and-events/news/2022/eninthespotlight/2022-10-27-new-en-17640-helps-evaluate-the-cybersecurity-of-ict-products/

## VII. CROSSCON Uses Cases

### UC1 - Device Multi-Factor Authentication

One of the main challenges of IoT devices is the access and authorization to the network or other specific resources. In recent years, Physically Unclonable Functions (PUFs) have been proposed as a solution for device authentication in constrained devices, mainly due to the scarcity of resources that hamper the utilization of regular cryptographic operations. However, PUF-based authentication has proven to be challenging to implement in practice and is vulnerable to a variety of attacks. CROSSCON proposes a multi-factor authentication (MFA) solution for IoT devices to improve their security and overcome the limitations of PUF-based authentication (Figure 2), providing a more robust defense against MITM (Man-in-the-Middle) attacks. The new authentication features are based on context and behavioral authentication, which may include both traditional authentication methods (PUF, private/public key scheme, and other credentials), and novel PUF-based authentication combined with environmental factors such as the network where the device is connected.

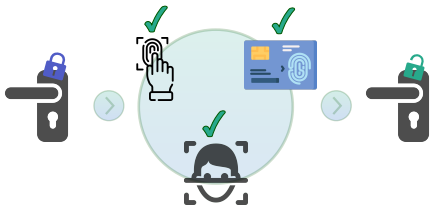

Fig. 2: Device Multi-Factor Authentication.

### UC2 - Firmware Updates of IoT Devices

It is very common to find IoT devices in the field without a secure firmware update system [5], [6]. Even those devices having firmware update mechanisms are in many cases not updated. An analysis performed over a total of 1.061.284 IoT devices shows the average age of the installed firmware is 19.2 months [16], leading to many vulnerabilities uncovered during large periods of time. Not being able to update IoT device firmware is one of the most common sources of vulnerability during the device lifecycle. Furthermore, an insecure update process also presents a major issue as it allows an attacker to upload malicious logic on the device. Updates and security patches can be digitally signed, so their integrity and authenticity can be verified. However, despite digital signatures, the problem of secure updates still persists, since: (i) updates often come as a bundle of libraries developed by different parties, (ii) the signatures are not always issued by a mutually trusted certification authority; and (iii) digital signatures do not give any guarantee on the logic of the update. CROSSCON aims at addressing the device's firmware update problem by providing a secure partial and full firmware update solution (Figure 3).

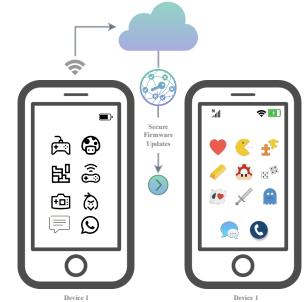

Fig. 3: Firmware Updates of IoT Devices.

### UC3 - Commissioning and Decommissioning of IoT devices

IoT Device Commissioning is the process by which connected devices acquire the necessary information and configuration parameters for their intended use or application, which can include security certificates, credentials, application configuration, and more (Figure 4). Commissioning is a critical step in the IoT device lifecycle, and it needs to happen before the device starts to perform its regular operation. As opposed, IoT Device Decommissioning is the process by which the commissioned information is removed from the device. The current solutions in the market, especially for resource-constrained devices, do not allow in many cases to generate unique random keys per device in a multi-stakeholder environment. This ends up in many cases with devices shipping with default and hard-coded credentials that can be exposed to attackers, which by stealing device secrets, can gain extra privileges within the device. CROSSCON is committed to implementing robust commissioning and decommissioning, ensuring the highest levels of security and reliability in IoT device operations.

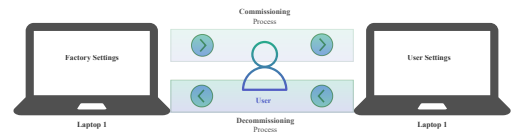

Fig. 4: Commissioning and Decommissioning of IoT devices.

### UC4 - Remote Attestation for Identification and Integrity Validation of Agricultural UAVs

Agricultural UAVs are essential for helping farmers in several tasks, e.g., seeding, fertilizing, irrigating, and pest controlling. Nevertheless, they also bring several privacy- and safety-related challenges. With a remote attestation feature, a method by which a client authenticates its hardware and software configuration to a remote host, we can ensure that a UAV runs a trusted software and hardware stack that meets the privacy, safety, and legal requirements (Figure 5).

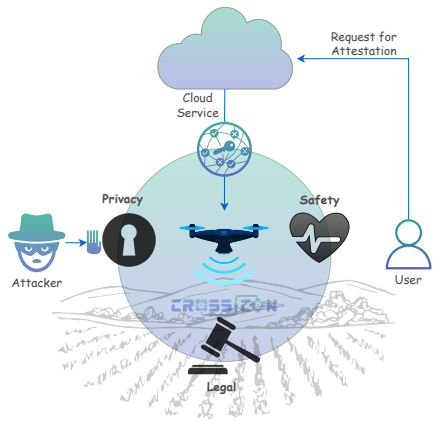

Fig. 5: Remote Attestation for Identification and Integrity Validation of Agricultural UAVs.

## UC5 - Intellectual Property Protection for Secure Multi-Tenancy on FPGA

Reconfigurable technology can be used for deploying compute-intensive tasks. To optimize its resource utilization, multiple tenants can share the reconfigurable platform. However, considering the security requirements of the CROSSCON stack, these resources must be temporal and/or spatially isolated. While the former ensures that one tenant has access to resources at a time, the latter provides access to different resources according to the tenant, supporting simultaneous tenants accessing the FPGA. This use case aims at providing secure multi-tenancy, ensuring that the workload of one tenant cannot interact with others nor affect the hardware resources, and that no data can be leaked by any means (Figure 6).

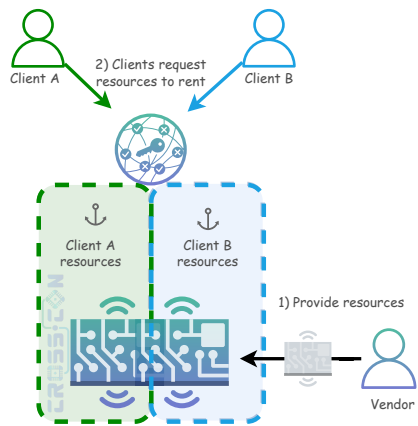

Fig. 6: Intellectual Property Protection for Secure Multi-Tenancy on FPGA.

## VIII. CONCLUSIONS

This white paper presents the CROSSCON project, a security stack that targets the current heterogeneity of the IoT landscape. It shows the current problems related to deploying secure IoT devices, the problem of their heterogeneity, and the regulations that are being put in place by the European Commission. It shows the security stack specifications, components, and the security features that are currently being developed, including the TEE abstraction and isolation, the CROSSCON Hypervisor, which is being already tested and supported on different devices and architectures. All the relevant information about the project status and on-going steps can always be found in the project's webpage (www.crosscon.eu), as well as in the public repository online (https://github.com/crosscon).